\documentclass{article}
\pdfoutput=1
\usepackage[utf8]{inputenc}
\usepackage{amsmath, amsthm, amssymb}
\usepackage{jinstpub}
\usepackage{graphicx}
\usepackage{support-caption}
\usepackage{subcaption}

\title{A novel approach to building micromegas detectors enabled by precision manufacturing}

\author[a,1]{J. Holmes\note{Corresponding Author}}
\author[b]{E. Galyaev}
\author[a]{R. Alarcon}
\author[b,c]{R. Acuna}
\author[a,b]{D. Blyth}
\author[a]{B. Fox}
\author[a,b]{N. Mullins}
\author[a,b]{K. Scheuer}

\affiliation[a]{Department of Physics, Arizona State University, Tempe, AZ 85287-1504, United States}
\affiliation[b]{Radiation Detection and Imaging Technologies, LLC, Tempe, AZ 85281, United States}
\affiliation[c]{School of Electrical, Computer and Energy Engineering, Arizona State University, Tempe, AZ 85287-5706, United States}


\abstract{Micromegas detectors are a relatively modern design concept for micropattern gas detectors, designed to handle high particle flux while providing high gain, high spatial resolution, and fast response times for a variety of radiation detection applications.  Due to the advancement of industry, building a micromegas detector without advanced in-house capabilities is now possible.  In this work, we present an innovative method to build micromegas detectors utilizing precision manufacturers to fabricate the core components.  Two detectors were built using the newly described method and are experimentally validated.}

\emailAdd{jmholmes@asu.edu}

\begin{document}

\maketitle



\section{Introduction}

The micromegas (micro-mesh-gaseous structure) detector, as first described by Giomataris et al. \cite{y._giomataris_micromegas:_1996} was initially designed to mitigate the space charge effect of ions generated in gaseous detectors with intrinsic electron multiplication.  The micromegas detector consists of two regions, a drift region where the primary particle ionizes the gas creating electron-ion pairs, and a very thin ($\sim$100~$\mu$m) amplification region, where the drifting electrons are multiplied.  Depending on the primary particle type, achievable gain typically ranges between 10$^3$ - 10$^5$.

The drift region and amplification region are separated by a micromesh, which allows drifting electrons to pass from the drift region to the amplification region.  The electric field strength in the drift region is defined by the cathode bias and micromesh bias while the electric field strength in the amplification region is defined by the micromesh bias and each anode bias.  By tuning the bias at the cathode, micromesh, and anodes, the electrons are able to pass through the micromesh with high efficiency.  Due to the kinematics of ions, only a small amount of ions are able to pass through the micromesh.  As a result, the micromegas detector quickly removes the ions generated from electron avalanche, thus preventing the ions from effecting the drift region fields.  The micromesh is a critical component of the micromegas detector and also the main source of complexity.  The ideal amplification region boundary has been described as being transparent to electrons, opaque to ions, perfectly permeable to the gas, and infinitely thin and flat (no spatial influence) \cite{kuger_micromesh-selection_2016}.

The vast majority of micromegas detectors are produced by CERN (European Organization for Nuclear Research) and CEA Saclay (Saclay Nuclear Research Centre) using the so called bulk micromegas \cite{giomataris_micromegas_2006} and microbulk micromegas \cite{andriamonje_development_2010} techniques.  To this point, researchers looking to implement micromegas detectors have to go through either CERN or CEA Saclay.  However, as precision manufacturing \cite{dornfeld_precision_2008} has improved over time, precision manufacturers are now able to produce the core micromegas components.  By outsourcing the fabrication of the most difficult components, researchers can now build their own micromegas detectors without advanced in-house capabilities.  This work outlines a new method for building micromegas detectors using parts fabricated by precision manufacturers followed by an experimental validation of the detectors.  Our goal is to improve accessibility to micromegas detectors by enabling researchers to build their own.

This research was conducted in part to develop a particle tracking sub-detector to improve beam entry position and orientation resolution of the Texas Active Target (TexAT) instrument at the Cyclotron Institute of Texas A\&M \cite{koshchiy_texas_2020}.

\section{Fabrication and assembly}
\begin{figure}
    \captionsetup[subfigure]{labelformat=empty}
    \centering
    \begin{subfigure}[b]{0.8\linewidth}
        \includegraphics[trim=0cm 0.5cm 0cm 2cm,width=\linewidth]{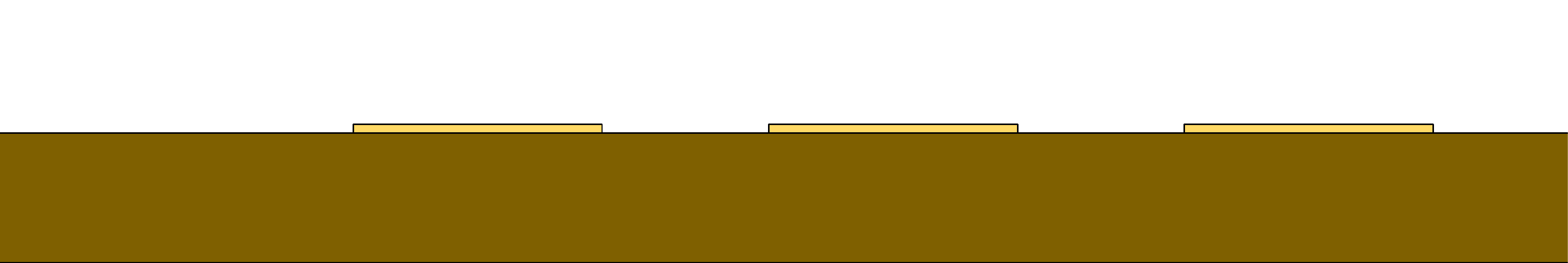}
        \caption{readout pattern and substrate}
    \end{subfigure}

    \begin{subfigure}[b]{0.8\linewidth}
        \includegraphics[trim=0cm 0.5cm 0cm 1cm,width=\linewidth]{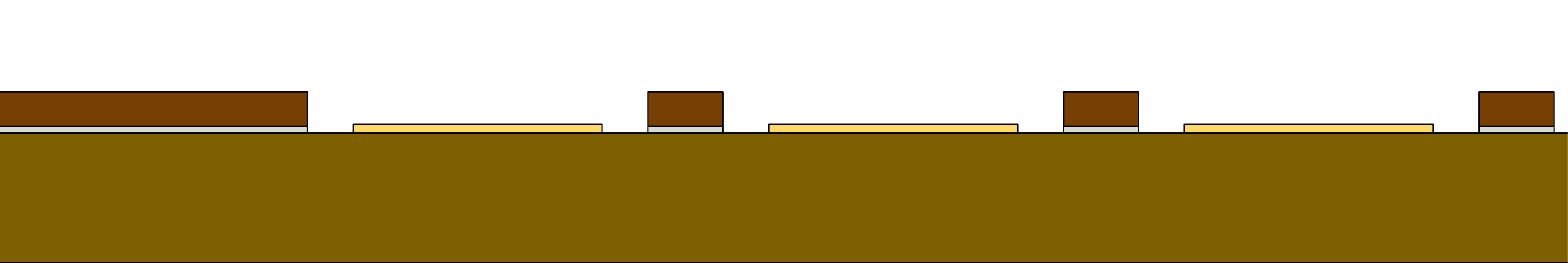}
        \caption{+ conforming dielectric film with adhesive}
    \end{subfigure}

    \begin{subfigure}[b]{0.8\linewidth}
        \includegraphics[trim=0cm 0.5cm 0cm -0.25cm,width=\linewidth]{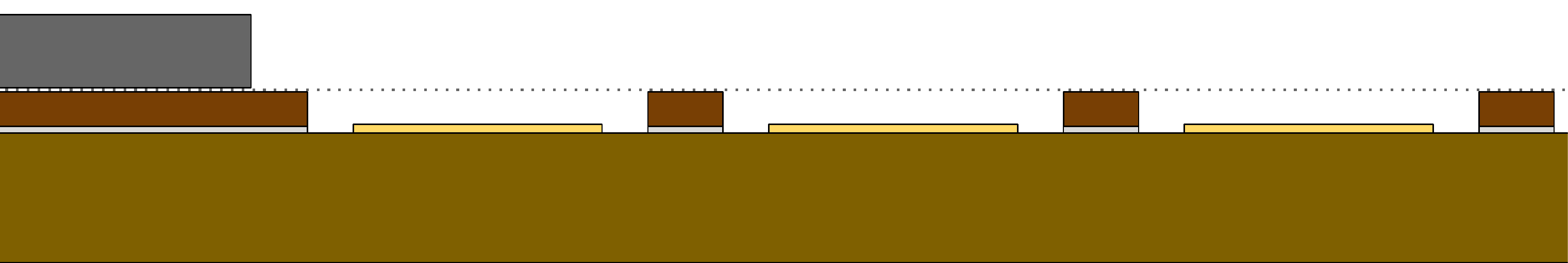}
        \caption{+ micromesh assembly}
    \end{subfigure}

    \caption{A diagram of the micromegas stack-up.  Each image identifies successive assembly steps and the resulting stack-up after each step.}
    \label{fig:mm_stackup_side}
\end{figure}

\begin{figure}[htpb]
    \captionsetup[subfigure]{labelformat=empty}
    \centering
    \begin{subfigure}[b]{0.384862\linewidth}
        \centering
        \includegraphics[trim=0cm 3cm 0cm -2cm,width=\linewidth]{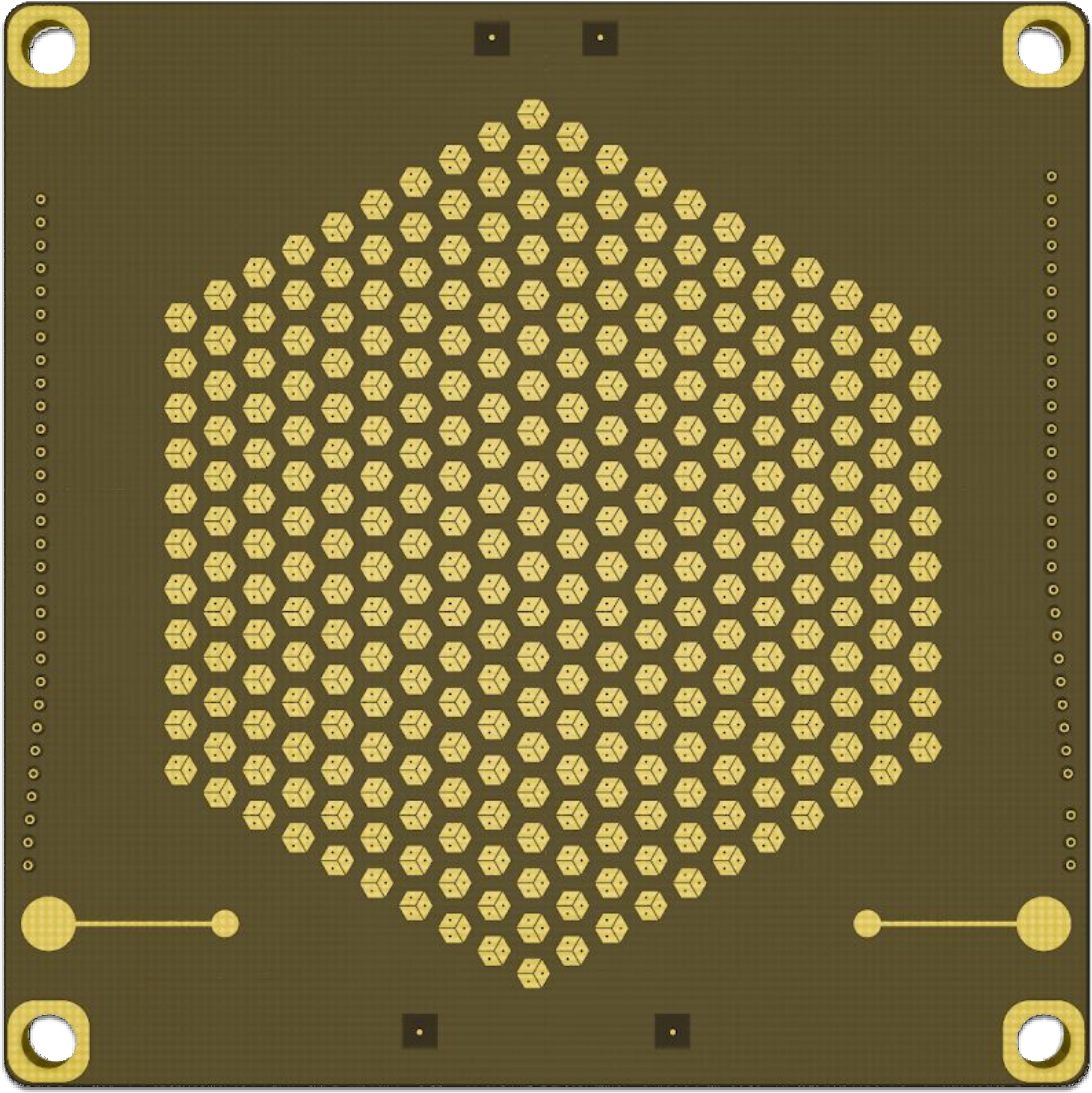}
        \caption{bare printed circuit board}
    \end{subfigure}
    
    \begin{subfigure}[b]{0.384862\linewidth}
        \centering
        \includegraphics[trim=0cm 3cm 0cm -2cm,width=\linewidth]{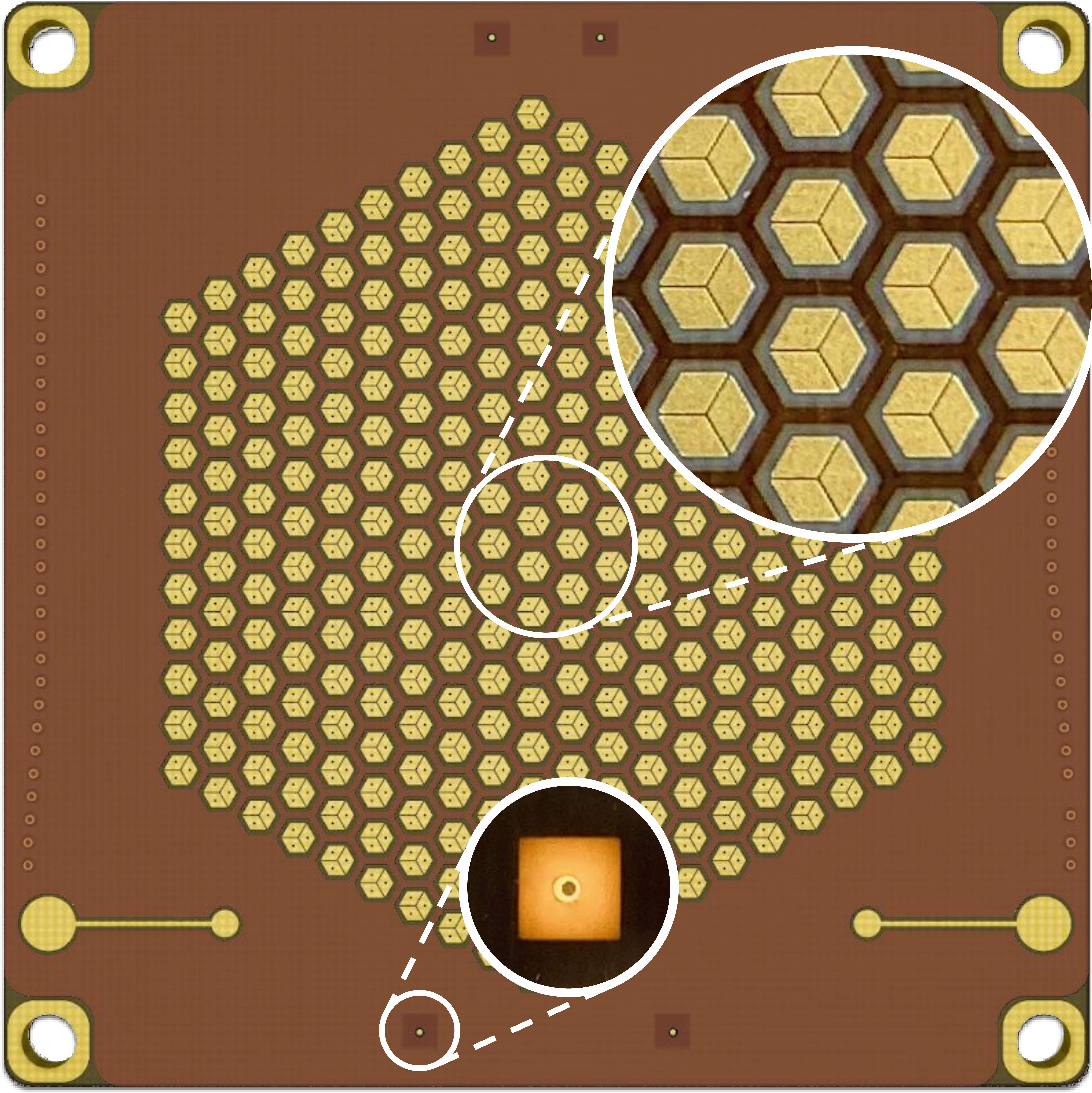}
        \caption{+ Kapton film (alignment pad is back-lit)}
    \end{subfigure}
    
    \begin{subfigure}[b]{0.384862\linewidth}
        \centering
        \includegraphics[trim=0cm 3cm 0cm -2cm,width=\linewidth]{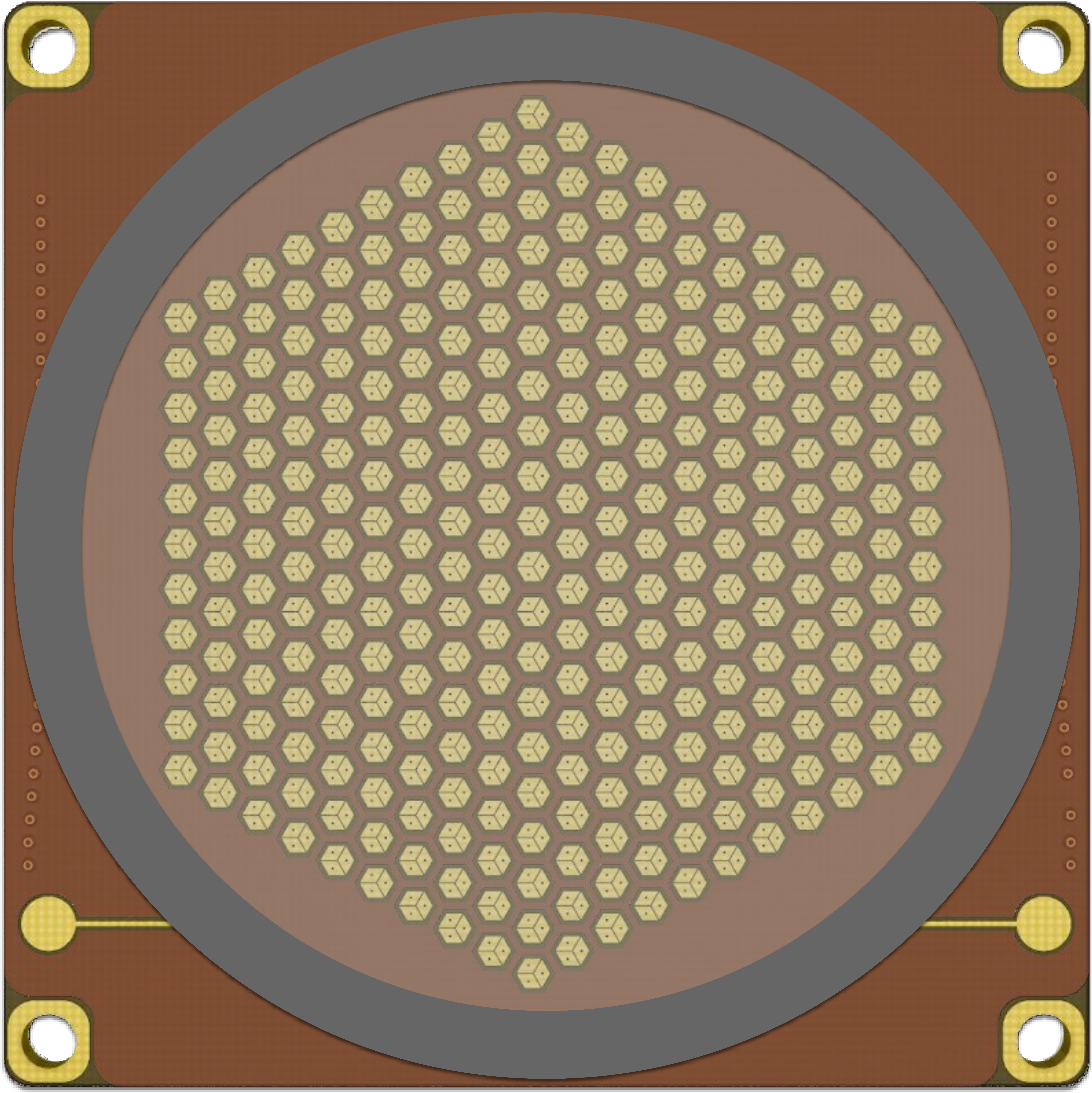}
        \caption{+ electroformed micromesh assembly}
    \end{subfigure}
    
    \caption{The micromegas designed for the TexAT instrument modification (1:1 scale).  Each image identifies successive assembly steps.}
    \label{fig:mm_stackup_top}
\end{figure}

Before detailing each step of the fabrication and assembly process, an overview of the process is given.  The most difficult and fundamental aspect of building a micromegas is placing the micromesh at a well-defined height above the readout pads.  To hold the micromesh above the readout pads, a thin dielectric film which conforms to (does not interfere with) the readout pattern is adhered to the substrate containing the pads, similar to applying a sticker.  The dielectric film acts as the standoff for the micromesh, defining its height above the readout pads.  A micromesh assembly is then held flush against the dielectric film.  The combined thickness of the dielectric film and adhesive minus the height of the readout pads defines the micromesh height.  The process as described requires three distinct parts to be fabricated by precision manufacturers: (1) the readout pattern and substrate, (2) the conforming dielectric film, and (3) the micromesh assembly.  For each assembly step, Figure \ref{fig:mm_stackup_side} shows a diagram of the micromegas stack-up while Figure \ref{fig:mm_stackup_top} shows a top view of the design intended for TexAT.  

\paragraph{A note on the assembly environment}
The assembly should be performed in a relatively low-dust environment.  If a clean room is not accessible, establishing a suitable environment for assembly may be feasible through good practices such as storing parts in new plastic bags, working in a carpet-free room that has been thoroughly cleaned, and maintaining a clean partition of the room.  It is advised to blow compressed gas or "canned air" over the parts and detector at each step to remove dust and debris no matter how clean the environment may be.  Visual inspection should be performed often as it can aid in identifying and removing debris.  The remainder of this section will detail the fabrication methods and assembly process.

\subsection{Readout pattern and substrate}
The readout pattern and substrate is designed using printed circuit board (PCB) layout software and fabricated using standard PCB processes.  The capabilities of PCB manufacturers have consistently improved over time in terms of minimum feature sizes, substrate materials, pad surface finishes, layer counts, via counts, and cost.  Regardless of the readout pattern, the readout pads on the top copper layer extrude above the PCB substrate surface by typically 18~$\mu$m (0.5 oz copper) or 36~$\mu$m (1 oz copper).  For an array of readout pads, internal connections should be made using via-in-pad, where the vias are filled with epoxy before the entire pad is coated with electroless nickel immersion gold (ENIG) or a different coating process.  The end result is a nearly featureless readout pad, which is important for minimizing the likelihood of sparking.  The readout pads should be arranged such that there is some room for the dielectric film between either single pads or groups of pads.  Figure \ref{fig:mm_stackup_top} shows the design for TexAT which has readout pads grouped in clusters of three allowing room for the dielectric film.

\subsection{Conforming dielectric film with adhesive}
Kapton is a good choice for the dielectric material due to its radiation tolerance \cite{garner_irradiation_1987, plis_review_2019} and its availability.  Kapton backed with an adhesive and release liner can be conformed to the readout pattern using laser-cutting.  Many companies offer laser-cutting of Kapton and will produce the part to specification.  Alternatively, the part can also be ordered through many PCB manufacturers as a stencil.  Our experience indicates that the minimum width within the capabilities of laser-cutting of Kapton is between 300 to 400~microns for films with thickness between 40 to 100 microns.  Therefore, using a conforming dielectric film to enforce the micromesh height results in a minimum spacing between groups of pads of about 300 to 400 microns plus a tolerance.  Notice the conservative tolerance between the Kapton film and the readout pattern in Figure \ref{fig:mm_stackup_top}.  The tolerance reduces the risk of interference.

\paragraph{Adhering the dielectric film to the PCB}
Adhering the dielectric film to the PCB presents two challenges: (1) alignment and (2) preventing bubbles from forming between the dielectric film and the PCB.  To attach the dielectric film, we use the so called "wet method" (often used in 3D printing).  Highly pure ASTM Type II deionized water is poured onto the PCB covering the entire board.  Upon setting the dielectric film onto the water, the water fills the space between the substrate and the dielectric film, leaving very few bubbles, while simultaneously holding the film securely in place with surface tension.  The water layer prevents the adhesive from making contact with the PCB and so the film can be aligned by hand into place without sticking.  Alignment is aided by markers designed into the PCB and the dielectric film.  In the area surrounding the alignment markers on the PCB, there is no copper on any layer, allowing for back-lighting.  When back-lit, the light is able to pass through the PCB and is blocked by the copper alignment marker (a pad on the PCB), creating an obvious visual aid.  The dielectric film contains a hole slightly larger than the alignment marker where the alignment marker should be aligned.  The dielectric film is guided by hand until each set of alignment markers is aligned.  See the back-lit alignment marker in the middle image of Figure \ref{fig:mm_stackup_top}.  Once the dielectric film is well aligned, excess water should be soaked with lint-free wipes.  The remaining water, including that which is beneath that dielectric film, should be left to evaporate.  Because the water is of high purity, no residue is left behind.  This alignment method is accurate to within about 100 to 200~microns.  Therefore, the tolerance should be at least 100 to 200~microns.

\subsection{Micromesh assembly}
The micromesh assembly is based on the electroforming process \cite{noauthor_precision_2020}.  At the limits when compared with steel woven mesh, electroformed mesh is 3-4 times thinner, has a higher line density, and a higher optical transparency, making the electroformed mesh closer to ideal than the steel woven mesh \cite{kuger_micromesh-selection_2016}.  We have chosen an electroformed nickel mesh with 5~$\mu$m thickness, 59\% optical transparency at 51~$\mu$m pitch (39~$\mu$m opening), however many other sizes are available.  This configuration can be made with areas of up to at least 28~cm x 28~cm, making it suitable for small to mid-sized micromegas detectors.  The mesh is produced using electro-deposition which is subsequently electro-bonded to a thin steel outer frame \cite{noauthor_precision_2020}, forming the micromesh assembly (see bottom image of Figure \ref{fig:mm_stackup_top}).  This precision manufacturing capability, producing a high quality micromesh assembly as an orderable part, was a primary factor in our choice of micromesh.

\begin{figure}
    \begin{center}
        \includegraphics[trim=0cm 0cm 0cm 0cm,clip=false,width=0.4\textwidth]{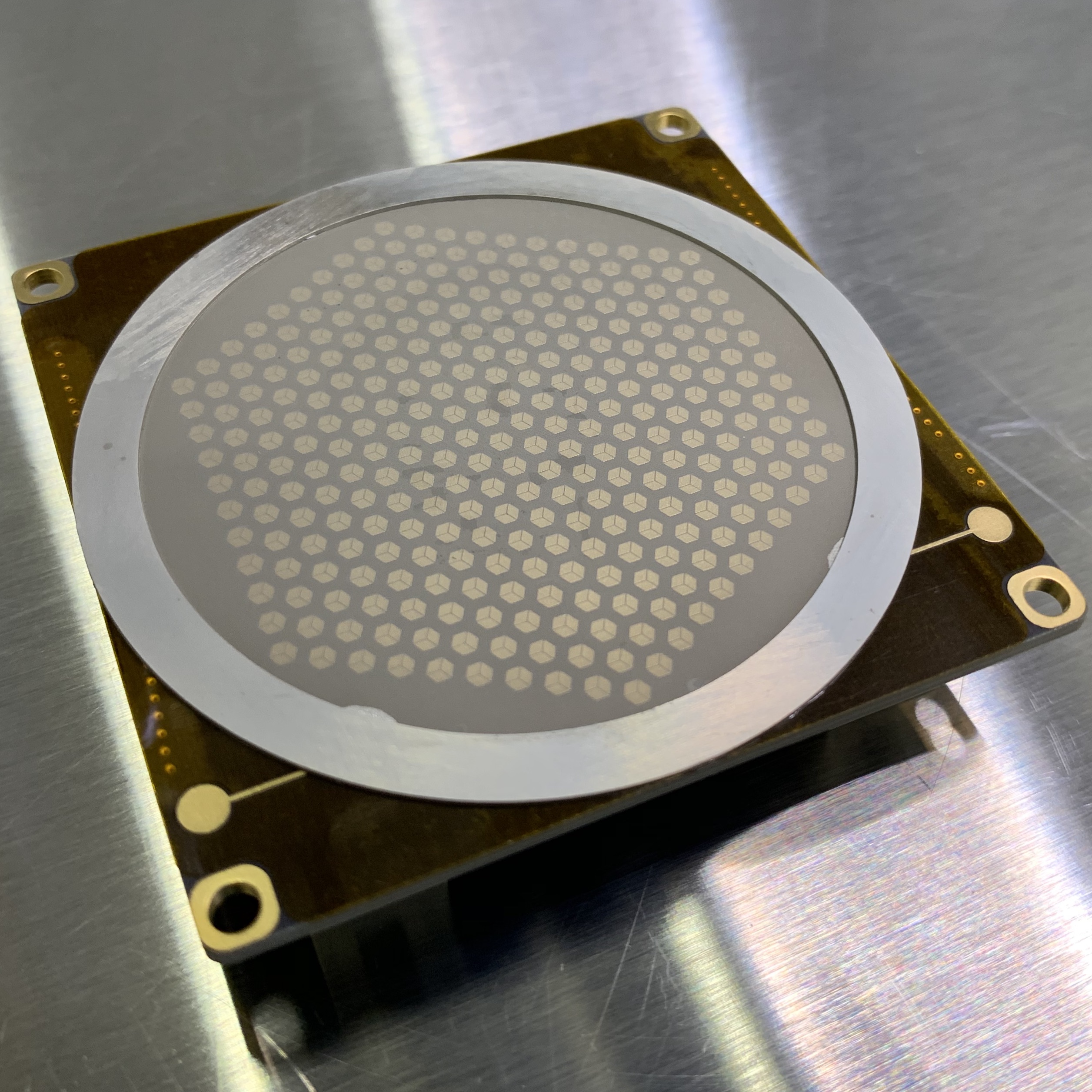}
        \caption{A micromegas detector produced using the methods described in this work.}
        \label{fig:fab}
    \end{center}
\end{figure}

The frame enforces the area of the mesh, prevents sag or wrinkling, and simplifies assembly.  The frame is not intended to enforce the mesh height and if it is too thick, the mesh may not rest flat on the dielectric film when pressure is applied.  Ideally, the electrostatic force when biased would be the only pressure applied to the mesh causing it to be flush with the dielectric film.  However, we attach the frame onto the dielectric film using a thin layer of low-outgassing epoxy (AA-BOND 2116) near the edges of the frame.  A bead of conducting silver epoxy is used to form an electrical contact between the mesh and the mesh-bias pad on the PCB.  A weight is laid over the frame while the epoxies cure.  An image of the detector fully assembled is shown in Figure \ref{fig:fab}.

\paragraph{A note on adhesive and epoxy outgassing}
To reduce the outgassing of adhesives and epoxy used in our designs, we minimized the exposed surface area of adhesives and epoxy as much as possible.  In a study examining reducing outgassing of epoxy resins, Gupta et al. \cite{gupta_outgassing_1977} found that covering epoxy with films having a low affinity for water vapor reduced the outgassing by a factor of 2-20 depending on the film material.

\section{Experimental validation and simulation}
\label{sec:exp}

As an electron traverses through a gas, the number of electrons created per unit path length or first Townsend coefficient, $\alpha$, depends on the field strength, gas concentration, and gas properties \cite{sharma_first_1993, urquijo_measurement_1999, giomataris_development_1998}.  The electron gain, $M$, relates to $\alpha$ by

\begin{equation}
    \label{eqn:alpha}
    ln(M) = \int^b_a \alpha \, dl \,\, ,
\end{equation}

\noindent where $a$ and $b$ are the electrode positions \cite{krajcar_bronic_townsend_2000}.  Assuming the field strength to be constant in the amplification region of a micromegas detector, the electron gain or multiplication factor, $M$, simplifies to

\begin{equation}
    \label{eqn:gain}
    M = e^{\alpha h} \,\, ,
\end{equation}

\noindent where $h$, equal to $b - a$, is the height of the amplification region \cite{giomataris_development_1998}.  To validate the new micromegas fabrication and assembly technique, the density-normalized first Townsend coefficient, $\alpha/N$, was found by measuring the gain over a range of density-normalized electric field strengths, $E/N$, for two micromesh heights, where $N$ is the gas concentration.  The results are compared with measurements made by Urquijo et al. \cite{urquijo_measurement_1999}, where data were collected for a parallel plate geometry.

\subsection{Setup}
\begin{figure}
    \begin{center}
        \includegraphics[trim=0cm 0cm 0cm 0cm,clip=false,width=\textwidth]{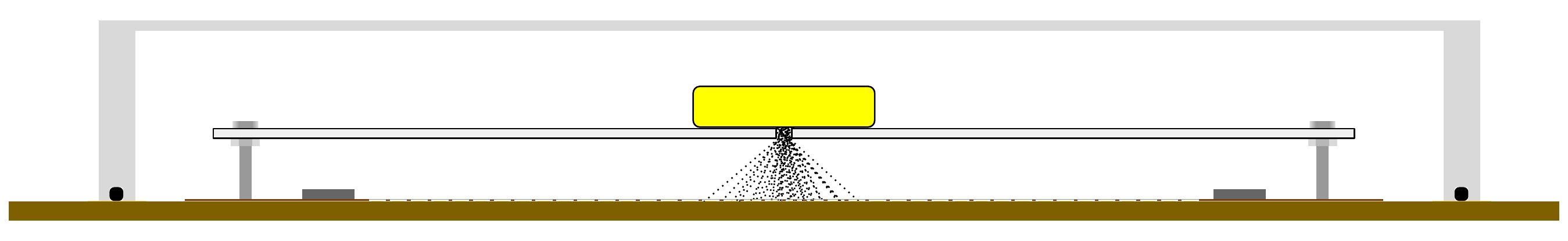}
        \caption{The gain measurement setup with a Po-210 alpha-particle source.  The cathode height is 5~mm.  This detector was fabricated using the methods described in this work with a very similar readout geometry as Figure \ref{fig:mm_stackup_top}, however it has a much larger active area of about 100~cm$^2$ and was designed to have the chamber mount directly to the PCB outside of the active area, sealed by a rubber gasket as shown.}
        \label{fig:setup}
    \end{center}
\end{figure}

A Po-210 source emitting 5.4~MeV alpha-particles sits on an aluminum plate with a small hole allowing alpha-particles to enter the drift region.  The plate acts as the cathode defining the drift region of the detector, sitting 5~mm above the micromesh.  The detector sits in a chamber containing P10, Argon with 10\% Methane (CH$_4$), at 1 atmosphere, between 293-300 Kelvin.  Figure \ref{fig:setup} shows a diagram of the setup geometry.  The alpha-particle flux is approximately 1~kHz/mm$^2$ over an area of 500~mm$^2$.  The use of alpha-particles to validate the detector was motivated by its potential use in the TexAT instrument at the Cyclotron Institute of Texas A\&M.

\begin{figure}
    \begin{center}
        \includegraphics[trim=5cm 1cm 0cm 0cm,clip=false,width=0.45\textwidth]{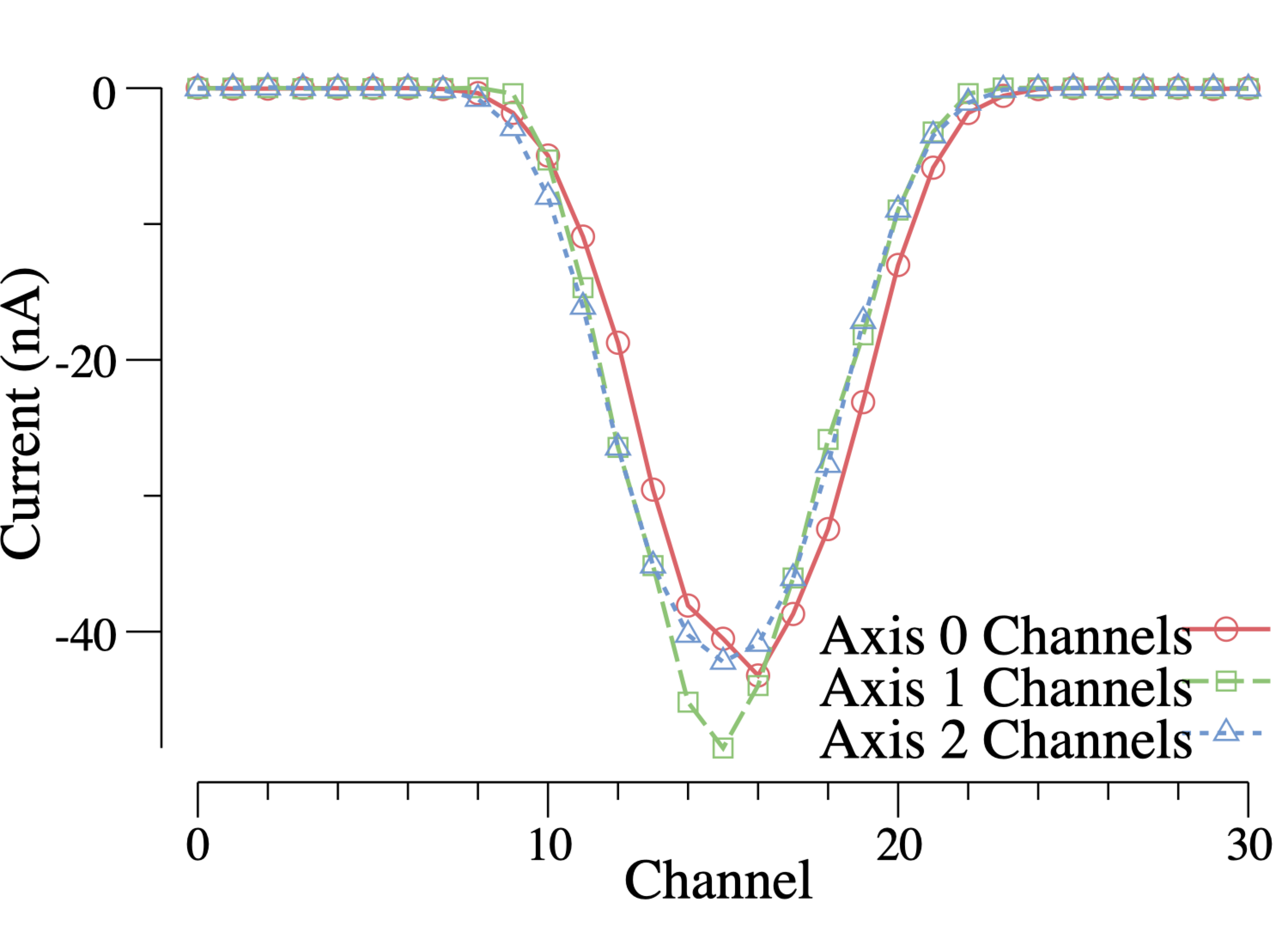}
        \caption{A sample of the detector response to a Po-210 alpha-particle source (gain on the order of 500).  The strip-like readout is connected such that the current is measured along three directions offset by 120$^{\circ}$, forming 3 projections of the detector current distribution.  This detector has 93 channels, 31 channels per projection.}
        \label{fig:source_proj}
    \end{center}
\end{figure}

\begin{figure}
    \begin{center}
        \includegraphics[trim=0cm 0cm 0cm 0cm,clip=false,width=0.5\textwidth]{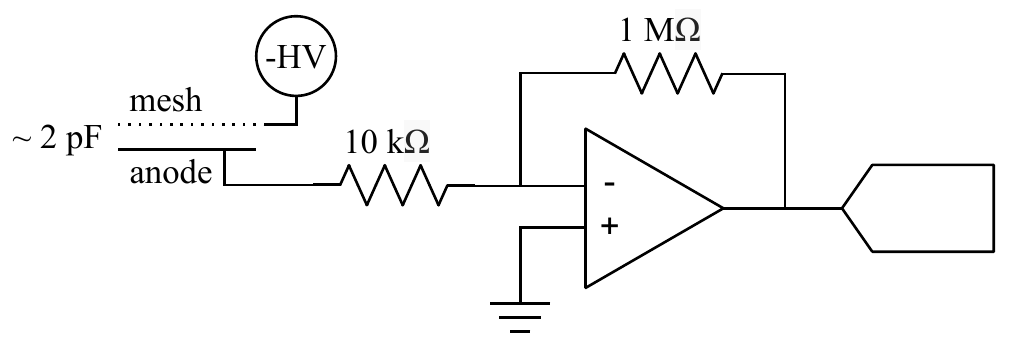}
        \caption{A schematic of the detector readout for a single channel.}
        \label{fig:schem}
    \end{center}
\end{figure}

Two detectors were fabricated using the methods described in this work, one having a micromesh height of 66~$\mu$m and the other with a height of 91~$\mu$m.  The detector has a unique strip-like readout, connected such that the detector current is measured along three directions offset by 120$^{\circ}$.  Figure \ref{fig:source_proj} shows a sample measurement of the current through each channel and the resulting three projections.  Transimpedance amplifiers (1~V/$\mu$A) with 10~kHz bandwidth sampled at 25~kHz measure the current through each anode channel.  Once digitized, the data are filtered to low bandwidth to reduce noise.  Each anode is held to pseudo-ground through a 10~k$\Omega$ resistor.  Each mesh-anode channel has a capacitance of about 2~pF.  Figure \ref{fig:schem} shows the a schematic of the readout electronics for each channel.  The 10~k$\Omega$ resistor protects the amplifier and detector by allowing the anode bias to vary depending on the current.  For example, in normal operation with 1~$\mu$A of current, the channel mesh-anode bias reduces by 10~mV, a negligible effect.  However, if a spark or discharge occurs resulting in 1~mA of current, the resistor will reduce the channel mesh-anode bias by 10~V momentarily, thereby reducing the gain and allowing the spark to be quenched.  The resistor has proven effective in protecting the micromesh and amplifier electronics.


\subsection{Measurement results}
\begin{figure}[htpb]
    \centering
    \begin{subfigure}[t]{0.4817987152\linewidth}
        \centering
        \includegraphics[trim=0cm 0cm 0cm 0cm,width=\linewidth]{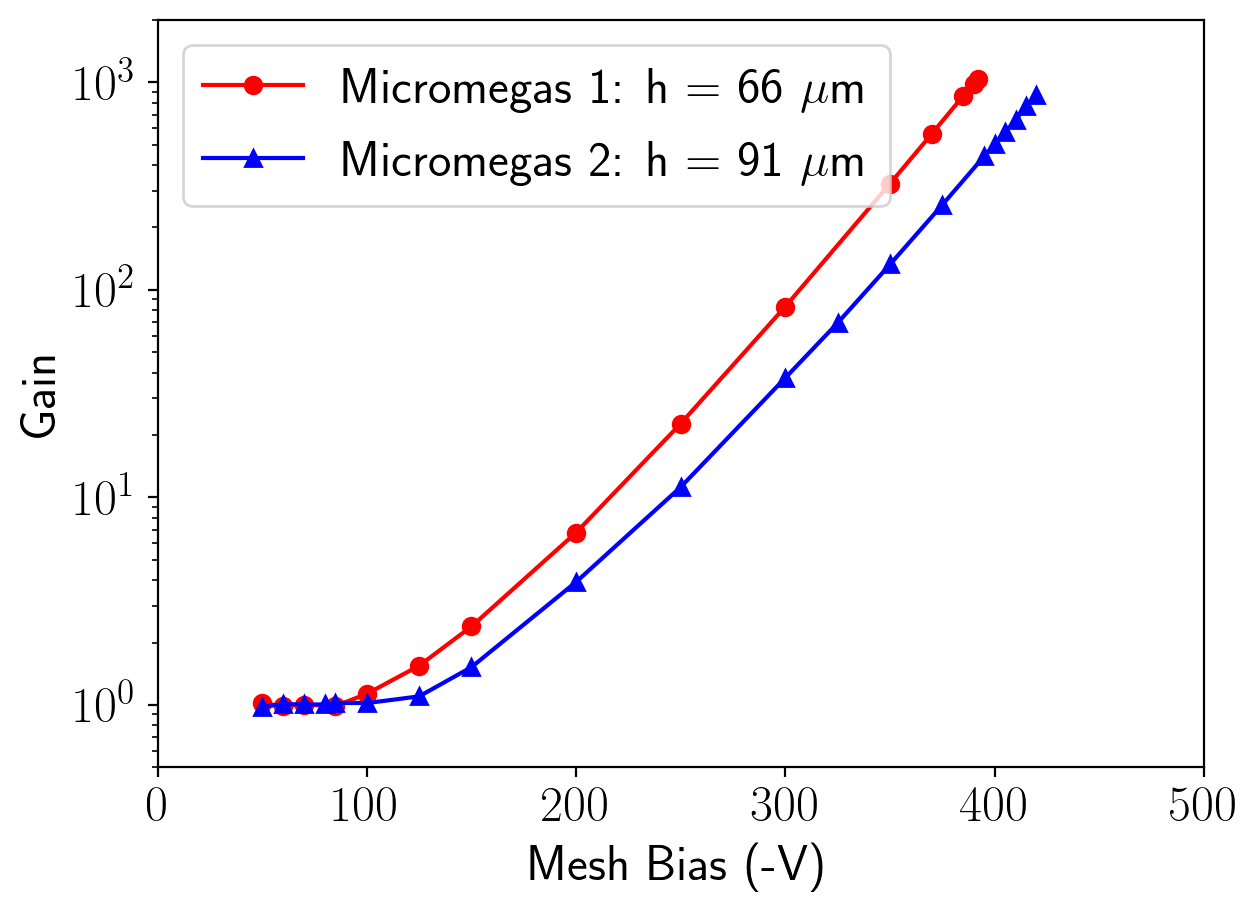}
        \caption{gain results}
        \label{fig:measurements_a}
    \end{subfigure}
    ~
    \begin{subfigure}[t]{0.5\linewidth}
        \centering
        \includegraphics[trim=0cm 0.19cm 0cm 0cm,width=\linewidth]{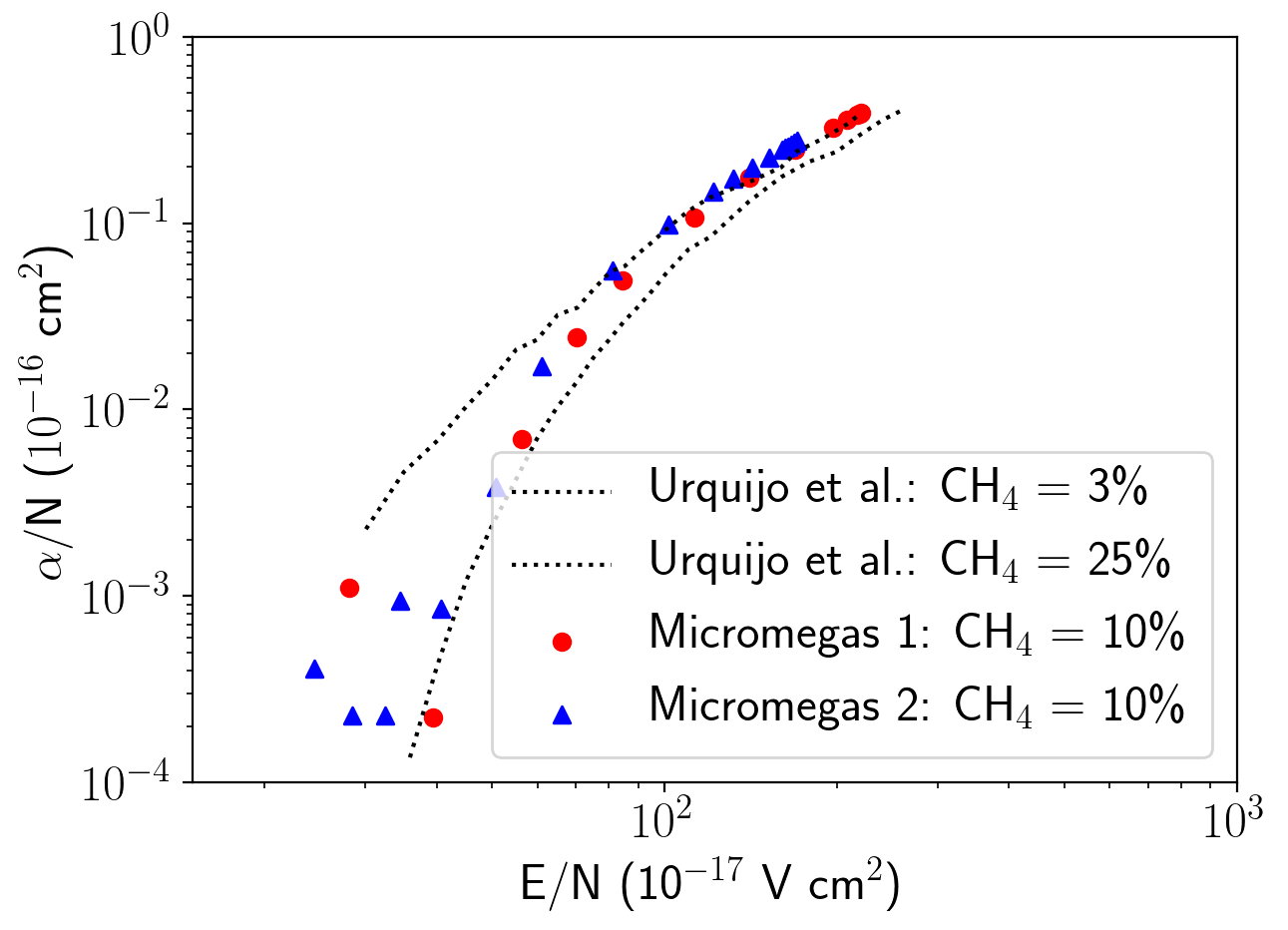}
        \caption{first townsend coefficient results}
        \label{fig:measurements_b}
    \end{subfigure}
    \caption{Measurement results for two detectors with different micromesh heights.  The first townsend coefficient results are compared to measurements published by Urquijo et al. \cite{urquijo_measurement_1999}, where parallel plates were used in argon-methane mixtures.}
    \label{fig:measurements}
\end{figure}

For each applied micromesh bias, each channel current is summed to get the total ionization current.  To ensure 100\% electron transparency of the micromesh for each applied micromesh bias, the cathode bias is varied until a maximum efficiency is reached (usually between 10-50~V/cm in the drift region).  At low micromesh bias, there is no multiplication; gain is 1.  As the micromesh bias is increased, the gain is initially plateaued at 1 until finally beginning to grow once the field strength reaches about 15~kV/cm.  The first 4-5 points along the plateau for each micromesh height are averaged to set the baseline gain of 1.  Gain measurements for both heights are shown in Figure \ref{fig:measurements_a}.  The maximum acheivable gain before catastrophic sparking (sparking rate in any channel at above $\sim$1~Hz) was measured to be near 10$^3$ for both detectors.  Sparking occurs when the Raether limit of about 10$^8$ electrons is approached \cite{peskov_study_2001}.  Because each alpha-particle is generating on the order of $10^4 - 10^5$ electron-ion pairs in the drift region, catastrophic sparking is expected as the gain approaches 10$^3$.  This was consistent with our results and explains why both detectors were limited to gain approaching 10$^3$.

From the gain measurements, the first Townsend coefficients can be directly solved.  Normalizing the first Townsend coefficient by the gas concentration,

\begin{equation}
    \alpha/N = ln(M)/(h \, N) \,\, ,
\end{equation}

\noindent where $N = 2.69$ x 10$^{19}$ cm$^{-3}$ is the gas concentration (number density) at STP.  The first Townsend coefficient measurements for both micromesh heights shown in Figure \ref{fig:measurements_b} compare well with the measurements made by Urquijo et al. \cite{urquijo_measurement_1999} suggesting that the field strength spanning the amplification region for both detectors can be approximated by a constant value and effectively treated as parallel plates, as equation \ref{eqn:gain} assumes.  This approximation is expected to break down as the micromesh height approaches the line-pitch within the micromesh.

\paragraph{Uncertainty estimates}
There are two dominant sources of uncertainty, both systematic, in the gain measurements.  The error in the baseline (gain = 1) measurement is about 1.5\% directly propagating into the gain measurements.  The greatest source of error is due to uncertainty in the micromesh height, between 3-5~$\mu$m, which propagates to about 10-15\% in the gain measurements.

\subsection{Electric fields simulation}

\begin{figure}[htpb]
    \centering
    \begin{subfigure}[t]{0.47\linewidth}
        \centering
        \includegraphics[trim=0cm 0cm -5cm 0cm,width=\linewidth]{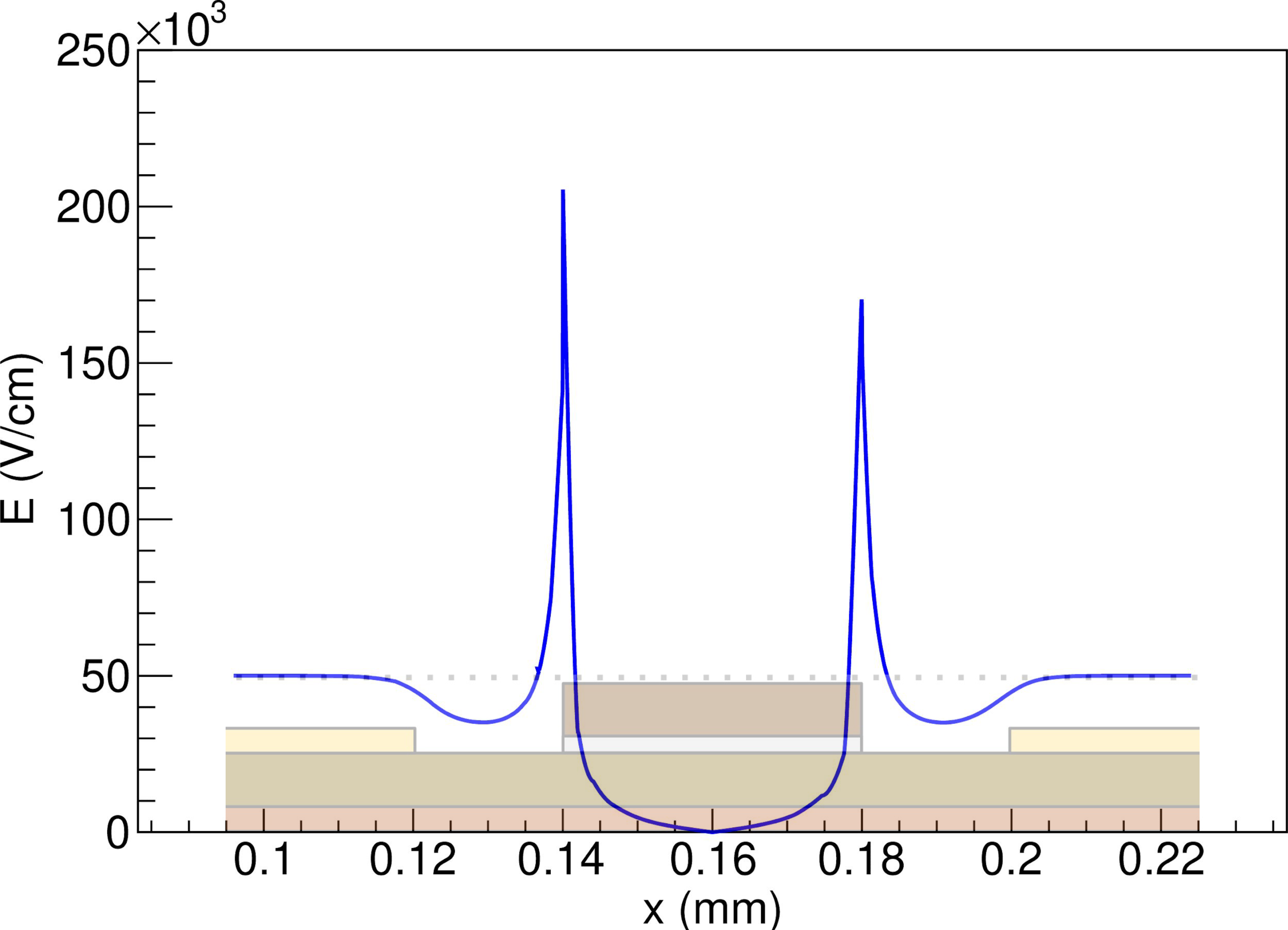}
        \caption{simulated electric field magnitude near the micromesh}
        \label{fig:evsx}
    \end{subfigure}
    ~
    \begin{subfigure}[t]{0.47\linewidth}
        \centering
        \includegraphics[trim=-5cm 0cm 0cm 0cm,width=\linewidth]{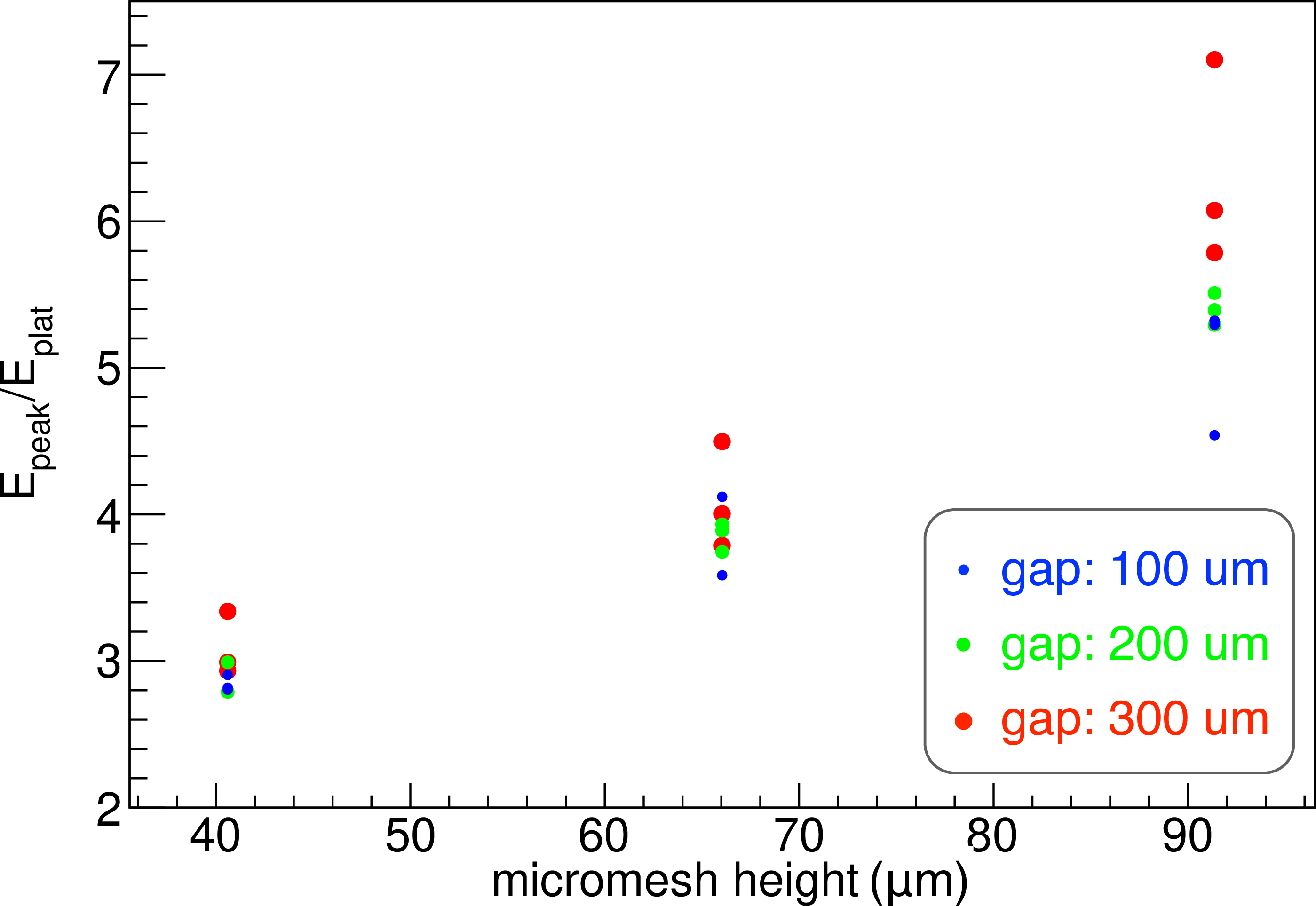}
        \caption{simulated electric field results}
        \label{fig:opera}
    \end{subfigure}
    \caption{The electric field simulation results for the conforming dielectric stand-off design.  Results are shown for varying geometric parameters.  The sharp peaks correspond with micromesh, dielectric, and gas triple-junctions.}
    \label{fig:fields}
\end{figure}

During the development stage of the detector, it was observed that sparking would sometimes occur where the micromesh, dielectric, and gas meet, known as a triple-junction.  The electric field strength can be greatly enhanced at a triple-junction \cite{tran_duy_partial_2008}.  To better understand geometric effects on the electric field strength near the triple-junction, the electric fields were simulated using the Opera finite element analysis software tool-set.  For one of the simulation configurations, Figure \ref{fig:evsx} shows the calculated field strength near the micromesh.  The sharp peaks in the field strength have the effect of limiting the maximum possible field strength above the readout pads since the peaks may exceed breakdown threshold.  Ideally, the ratio of peak field strength to plateau field strength (field strength above the readout pads) would be 1.  This ratio was calculated over a range of geometric configurations.  Although the actual mechanism for breakdown is complex, it is assumed that a lower peak field strength to plateau field strength ratio allows the detector to operate at higher field strengths before breakdown.

For the calculation, three geometric parameters were considered: the micromesh height above the readout pad, the gap between the dielectric walls and the readout pads, and the PCB thickness between the top layer and the first inner ground plane (PCB material: FR4).  For each configuration, the electric field magnitude along a line passing just below the micromesh and through the dielectric was averaged using six slightly different positions to reduce simulation noise.  In total, 27 configurations were simulated: micromesh heights \{40 $\mu$m, 65 $\mu$m, 90 $\mu$m\}, gap sizes \{100 $\mu$m, 200 $\mu$m, 300 $\mu$m\}, and FR4 thicknesses \{75 $\mu$m, 150 $\mu$m, 500 $\mu$m\}.  Figure \ref{fig:opera} shows the results for each configuration.  Of the three parameters which were varied, only the micromesh height was found to significantly effect the peak field strength to plateau field strength ratio.  In most detection scenarios, the goal is to configure the geometry such that the detector can achieve the highest gain before breakdown.  These results suggest that designs utilizing the conforming dielectric as the micromesh stand-off may be most suitable for detectors that are optimal at small micromesh heights.

\section{Conclusion}
Designing and building a micromegas detector without advanced in-house capabilities is now possible in part due to the advancement of industry.  The most difficult micromegas components to build, the micromesh assembly and the dielectric standoff to hold the micromesh at a well defined height, can be manufactured by precision manufacturers to meet the specifications required for many applications.

The detectors were characterized by measuring the response to alpha-particles emitted by Po-210.  The results were comparable to measurements made by Urquijo et al., suggesting that the amplification region above the readout pads is adequately modelled as an ideal micromegas detector (parallel plates) for heights above 66~$\mu$m.

Six micromegas detectors with various micromesh heights have been prepared to be tested at the Cyclotron Institute as an addition to the TexAT instrument.

\section{Acknowledgment}

This material is based upon work supported by the U.S. Department of Energy, Office of Science, Nuclear Physics program office under Award Number DE-SC0015136.

\paragraph{Disclaimer}
This report was prepared as an account of work sponsored by an agency of the United States Government. Neither the United States Government nor any agency thereof, nor any of their employees, makes any warranty, express or implied, or assumes any legal liability or responsibility for the accuracy, completeness, or usefulness of any information, apparatus, product, or process disclosed, or represents that its use would not infringe privately owned rights. Reference herein to any specific commercial product, process, or service by trade name, trademark, manufacturer, or otherwise does not necessarily constitute or imply its endorsement, recommendation, or favoring by the United States Government or any agency thereof. The views and opinions of authors expressed herein do not necessarily state or reflect those of the United States Government or any agency thereof.

\bibliographystyle{ieeetr}
\bibliography{Library}

\end{document}